\documentclass[twocolumn,aps,prl,floatfix,superscriptaddress,notitlepage]{revtex4-1}
\usepackage[utf8]{inputenc}
\usepackage[OT1]{fontenc}
\usepackage{amsmath}
\usepackage{amsfonts}
\usepackage{amssymb}
\usepackage{bm}
\usepackage{graphicx}
\usepackage[left=2cm,right=2cm,top=2cm,bottom=2cm]{geometry}
\usepackage{longtable,booktabs}
\allowdisplaybreaks

\usepackage{dcolumn}
\usepackage{siunitx}
\usepackage{multirow}

\usepackage{color}
\definecolor{BLUE}{rgb}{0.0,0.0,1.0}

\usepackage{soul}
\soulregister\cite7
\soulregister\ref7

\newcommand{\veps}{\varepsilon}
\newcommand{\balpha}{\bm{\alpha}}

\newcommand{\bp}{\bm{p}}

\newcommand{\be}{\begin{eqnarray}}
\newcommand{\ee}{\end{eqnarray}}

\newcommand{\matr}[3]{\langle #1 | #2 | #3 \rangle}
\newcommand{\bra}[1]{\langle#1|}
\newcommand{\ket}[1]{|#1\rangle}

\begin{document}

\title{\textit{Ab initio} calculations of energy levels in Be-like xenon: \\ strong interference between electron-correlation and QED effects}

%

\author{A.~V.~Malyshev}
\affiliation{Department of Physics, St.~Petersburg State University, Universitetskaya 7/9, 199034 St.~Petersburg, Russia  
\looseness=-1}

\author{D.~A.~Glazov}
\affiliation{Department of Physics, St.~Petersburg State University, Universitetskaya 7/9, 199034 St.~Petersburg, Russia  
\looseness=-1}

\author{Y.~S.~Kozhedub}
\affiliation{Department of Physics, St.~Petersburg State University, Universitetskaya 7/9, 199034 St.~Petersburg, Russia  
\looseness=-1}

\author{I.~S.~Anisimova}
\affiliation{Department of Physics, St.~Petersburg State University, Universitetskaya 7/9, 199034 St.~Petersburg, Russia  
\looseness=-1}

\author{M.~Y.~Kaygorodov}
\affiliation{Department of Physics, St.~Petersburg State University, Universitetskaya 7/9, 199034 St.~Petersburg, Russia  
\looseness=-1}

\author{V.~M.~Shabaev}
\affiliation{Department of Physics, St.~Petersburg State University, Universitetskaya 7/9, 199034 St.~Petersburg, Russia  
\looseness=-1}

\author{I.~I.~Tupitsyn}
\affiliation{Department of Physics, St.~Petersburg State University, Universitetskaya 7/9, 199034 St.~Petersburg, Russia  
\looseness=-1}


\begin{abstract}

The strong mixing of close levels with two valence electrons in Be-like xenon greatly complicates \textit{ab initio} QED calculations beyond the first-order approximation. Due to a strong interplay between the electron--electron correlation and QED effects, the standard single-level perturbative QED approach may fail, even if it takes into account the second-order screened QED diagrams. In the present Letter, the corresponding obstacles are overcome by working out the QED perturbation theory for quasidegenerate states. The contributions of all the Feynman diagrams up to the second order are taken into account. The many-electron QED effects are rigorously evaluated in the framework of the extended Furry picture to all orders in the nuclear-strength parameter $\alpha Z$. The higher-order electron-correlation effects are considered within the Breit approximation. The nuclear recoil effect is accounted for as well. The developed approach is applied to high-precision QED calculations of the ground and singly excited energy levels in Be-like xenon. The most accurate theoretical predictions for the binding and excitation energies are obtained. These results deviate from the most precise experimental value by $3\sigma$ but perfectly agree with a more recent measurement.

\end{abstract}


\maketitle



Highly charged ions serve as ideal laboratories for testing methods of bound-state quantum electrodynamics (QED) in the presence of strong electromagnetic fields. Nowadays, the most stringent tests are related to the Lamb-shift measurements in H- and Li-like uranium~\cite{Stoehlker:2000:3109, Gumberidze:2005:223001, Schweppe:1991:1434, Brandau:2003:073202, Beiersdorfer:2005:233003}, for associated theory see, e.g., Refs.~\cite{Yerokhin:2006:253004, Kozhedub:2008:032501, Sapirstein:2011:012504}. Nevertheless, these results are not sufficient for the comprehensive comparison of theory and experiment. In this respect, Be-like ions being the simplest example of atomic systems with more than one electron in a valence shell are of particular interest. QED effects in Be-like ions are of the same order of magnitude as in He- and Li-like ions (see, e.g., the recent review~\cite{Indelicato:2019:232001} and references therein). Therefore, these ions have the capacity to provide tests of QED at the level comparable to that of Li-like systems. However, in contrast to three-electron ions, essentially new many-particle QED effects appear in this case. Electromagnetic interaction of the $L$-shell electrons, combined with intershell correlation effects, leads to a strong mixing of the close energy levels of the same symmetry. As a result, highly charged Be-like ions pose a serious challenge to atomic structure calculations within the framework of bound-state QED. 
Despite numerous relativistic calculations of the energy levels in these systems~\cite{Armstrong:1976:1114, Cheng:1979:111, Lindroth:1992:2771, Zhu:1994:3818, Safronova:1996:4036, Chen:1997:166, Safronova:2000:1213, Majumder:2000:042508, Gu:2005:267, Ho:2006:022510, Cheng:2008:052504, Sampaio:2013:014015, Yerokhin:2014:022509, Wang:2015:16, El-Maaref:2015:2, Li:2017:720}, the QED effects have been included previously at best within some one-electron approximations only. Different theoretical approaches show significant scatter of the results when compared with the available experimental data~\cite{Edlen:1983:51, Moller:1989:333, Trabert:1990:860, Moller:1991:223, Buttner:1992:693, Beiersdorfer:1993:3939, Beiersdorfer:1995:2693, Trabert:2003:042501, Draganic:2003:183001, Feili:2005:48, Schippers:2012:012513, Bernhardt:2015:144008} providing strong motivation for rigorous QED calculations of Be-like ions. The many-electron QED effects for the ground state of Be-like ions were evaluated in our recent works~\cite{Malyshev:2014:062517, Malyshev:2015:012514}, where the calculations were performed using the QED perturbation theory for a single level. However, due to the proximity of all the $n=2$ levels, including the ground state, the accuracy of this evaluation remains unclear, unless the calculations based on \textit{ab initio} QED approach for quasidegenerate states are performed. The present work is intended to solve this long-standing and extremely difficult problem. The development of the rigorous QED theory for the high-precision calculations of the ground and singly excited states in Be-like ions is the primary goal of this study. To the best of our knowledge, the QED calculations of the quasidegenerate states at such a level have never been performed previously for the systems with more than two electrons. Moreover, the three-dimensional model subspace of quasidegenerate levels, which we employ in the present paper, is also considered for the first time in the framework of \textit{ab initio} approach. It is the application of such sophisticated methods that allows us to accurately address an issue of strong interference between electron--electron interaction and QED effects.

The natural zeroth-order approximation for the systematic QED description of highly charged ions is provided by the Dirac equation 
\begin{equation}
\label{eq:DirEq}
\left[ \balpha \cdot \bp + \beta m + V \right] \psi_n = \veps_n \psi_n \, .
\end{equation}
By substituting the Coulomb potential of the nucleus~$V_{\rm nucl}$ as the binding potential~$V$ in Eq.~(\ref{eq:DirEq}) one comes to the Furry picture of QED~\cite{Furry:1951:115}. The electron--nucleus interaction is taken into account to all orders in~$\alpha Z$ in this way ($\alpha$ is the fine-structure constant, $Z$ is the nuclear charge number). 
In order to partially account for the electron--electron interaction effects from the very beginning, the initial approximation can be modified by adding some local screening potential, $V=V_{\rm nucl}+V_{\rm scr}$. This corresponds to the extended version of the Furry picture~\cite{Sapirstein:2001:022502, Sapirstein:2001:032506, Sapirstein:2002:042501, Sapirstein:2003:022512, Chen:2006:042510, Sapirstein:2006:042513, Yerokhin:2007:062501, Artemyev:2007:173004, Kozhedub:2010:042513, Sapirstein:2011:012504, Artemyev:2013:032518, Malyshev:2017:022512}. The remaining part of the interelectronic interaction as well as the interaction with the quantized electromagnetic field are to be considered within appropriate perturbation theory~(PT). 

Standard nondegenerate PT suits well for single (isolated) levels, such as the ground state~$1s^2$ of He-like ions~\cite{Yerokhin:1997:361}, or when one studies, e.g., the fine-structure splitting~$2p_{j}$--$2s$ in Li-like ions~\cite{Kozhedub:2010:042513, Sapirstein:2011:012504}. 
However, considering more complicated systems or excited states one inevitably encounters close levels with the same symmetry which are mixed strongly by the electron--electron interaction. Application of the extended Furry picture allows one to lift the degeneracy in some cases. Nevertheless, the proper treatment of these systems requires employing PT for quasidegenerate levels. Both types of the QED perturbation series can be conveniently constructed in the framework of the two-time Green's function (TTGF) method~\cite{TTGF}. For a set of $s$ quasidegenerate levels, the TTGF method implies the evaluation of the $s \times s$ matrix~$H$ which acts in the model subspace~$\Omega$ spanned by the unperturbed wave functions of the states under consideration. The energies can be obtained by diagonalizing the matrix~$H$. PT for a single level corresponds to $s=1$. 

In the present work, we aim at the \textit{ab initio} evaluation of the ground and singly excited energy levels in Be-like xenon. 
Xenon is chosen as the object of study since it is a very suitable system to probe the nonperturbative (in $\alpha Z$) QED methods for evaluating the many-particle QED effects: on the one hand, the nuclear charge number $Z$ is high enough to make the QED contributions rather pronounced and, on the other hand, there must be a strong interference between the correlation and QED effects.
Despite the fact that our calculations are fully relativistic, we employ the $LS$-coupling notations. The excited states with the total angular momentum equal to $J=0$ and $J=2$, namely $2s2p \, ^3P_0$ and $2s2p \, ^3P_2$ (here and in what follows, the $K$ shell is omitted for brevity), are considered as the single levels. The corresponding unperturbed wave functions in the $jj$ coupling read as $(2s2p_{1/2})_0$ and $(2s2p_{3/2})_2$, respectively. 
The states $2s2p \, ^3P_1$ and $2s2p \, ^1P_1$ with $J=1$ are studied within the two approaches: (a) as the isolated levels, starting from the initial approximations $(2s2p_{1/2})_1$ and $(2s2p_{3/2})_1$; (b) as a pair of quasidegenerate levels within the two-dimensional subspace $\Omega$.
Finally, the binding energy of the ground $2s2s \, ^1S_0$ state is evaluated by means of three independent approaches:
(a) as the isolated $(2s2s)_0$ level; (b) as one of the two quasidegenerate levels, $(2s2s)_0$ and $(2p_{1/2}2p_{1/2})_0$, within the two-dimensional subspace $\Omega$; (c) as one of the three quasidegenerate levels, including the $(2p_{3/2}2p_{3/2})_0$ configuration, for which the mixing coefficient can be even larger than for $(2p_{1/2}2p_{1/2})_0$~\cite{Armstrong:1976:1114, Braun:1984:book:rus2eng}. The approach (a) for the ground state reproduces the one which we have used in Ref.~\cite{Malyshev:2014:062517}.
In the Coulomb potential, the unperturbed levels forming the quasidegenerate subspaces are split only by the nuclear-size and relativistic effects. 


Let us briefly formulate our approach. In order to derive all the relevant calculation formulas, we start with the formalism in which the closed $1s^2$ shell is regarded as belonging to the vacuum~\cite{TTGF}, and there are two $L$-shell ``valence'' electrons. The redefinition of the vacuum is carried out by changing the sign before $i0$ in the electron-propagator denominators corresponding to the closed shell, i.e., the standard Green's function~$G$ is replaced as follows:
\begin{equation}
\label{eq:redefinition}
G(\omega) \equiv \sum_n \frac{ \ket{n} \bra{n} }{ \omega - \veps_n + i\veps_n 0 } 
\,\rightarrow\,
\sum_n \frac{ \ket{n} \bra{n} }{ \omega - \veps_n + i\eta_n 0 } \, ,
\end{equation}
where $\eta_n = \veps_n - \veps_{\rm F}$, and $\veps_{\rm F}$ is the Fermi energy. The Fermi energy is chosen to be higher than the energy of the closed-shell electrons but lower than the valence-electron energy. In this formalism, the one- and two-loop one-electron Feynman diagrams include the contributions describing the interaction between one valence electron and the closed shell. These contributions can be separated by considering the difference with the standard definition of the vacuum. 
For instance, the first-order self-energy and vacuum-polarization diagrams for the valence state $|v\rangle$ lead to the one-photon exchange contribution
\begin{equation}
\label{eq:1ph}
\Delta E^{(1)}_{\rm int} = \sum_c \left[ I_{vcvc}(0) - I_{cvvc}(\veps_v-\veps_c) \right] \, ,
\end{equation}
where $I_{abcd}(\omega)=\matr{ab}{I(\omega)}{cd}$, $I(\omega)=e^2\alpha_1^\mu\alpha_2^\nu D_{\mu\nu}(\omega)$, $\alpha^{\mu}=(1,\balpha)$, and $D_{\mu\nu}(\omega)$ is the photon propagator. We stress that the treatment of the one-electron diagrams does not depend on the type of PT. By employing this formalism, we obtain the well-known expressions for the intershell-interaction corrections derived previously for Li-like systems, see, e.g., Refs.~\cite{Yerokhin:1999:3522, Artemyev:1999:45, Yerokhin:2001:032109, Sapirstein:2011:012504}. 

Now we turn to the discussion of the two-electron diagrams within the formalism with the redefined vacuum. These diagrams contain the three-electron contributions corresponding to the interaction of both the valence electrons with the $1s^2$~core. In case of He-like ions and two-dimensional subspace~$\Omega$, the formal expressions for the second-order two-electron contributions were derived within the TTGF method in Refs.~\cite{LeBigot:2001:040501_R, Artemyev:2000:022116, Artemyev:2005:062104}, see also Ref.~\cite{Kozhedub:2019:062506}. In the present work, we have generalized these expressions to deal with an arbitrary number of quasidegenerate levels. The most complicated and time-consuming is the derivation of the formulas for the two-photon exchange diagrams~\cite{ Blundell:1993:2615, Shabaev:1994:4489, Lindgren:1995:1167, Mohr:2000:052501}.
We restrict our consideration only to the contribution of the ladder diagram. 
This contribution is naturally divided into the reducible (``red'') and irreducible (``irr'')  parts. The reducible part involves the terms with the intermediate states coinciding with the quasidegenerate levels under consideration, while the irreducible part corresponds to the remainder. The reducible part is not affected by the redefinition of the vacuum, and its contribution to~$H$, arising from the interaction between the valence electrons, reads as
\begin{widetext}
\begin{align}
\label{eq:H2ik_ld_red}
H^{\rm red}_{ik} = -\frac{ 1 }{ 2 } \, \sum_{P} (-1)^P 
   \!\!\sum_{n_1n_2}^{E_n^{(0)} = E_1^{(0)} \ldots E_s^{(0)}}\!\!
   \frac{i}{2\pi} \int_{-\infty}^\infty \! d\omega \,
&
 \Bigg[\,
   \frac{ I_{Pi_1Pi_2n_1n_2}(\omega-\veps_{Pi_1}) I_{n_1n_2k_1k_2}(\veps_{k_1}-\omega) }
        { \big( \omega - \veps_{n_1} - i0 \big) \big( \omega +  \veps_{n_2} - \bar{E}_{ik}^{(0)} - i0 \big) } 
 + \{ 1 \leftrightarrow 2 \}
 \,\Bigg] \, ,
\end{align}
where the indices $i$ and $k$ enumerate the quasidegenerate states, $P$ is the permutation operator, $E_n^{(0)}=\veps_{n_1}+\veps_{n_2}$, $\bar{E}_{ik}^{(0)} = (E_i^{(0)}+E_k^{(0)})/2$, and $\{ 1 \leftrightarrow 2 \}$ means the expression with the transposed indices 1 and 2. 
The contribution of the irreducible part of the ladder diagram within the employed formalism can be expressed as follows:
\begin{align}
\label{eq:H2ik_ld_irr}
\tilde{H}^{\rm irr}_{ik} = \frac{ 1 }{ 2 } \, \sum_{P} (-1)^P 
   \!\!\sum_{n_1n_2}^{E_n^{(0)} \neq E_1^{(0)} \ldots E_s^{(0)}}\!\! 		
   \frac{i}{2\pi} \int_{-\infty}^\infty \! d\omega \,
&
 \Bigg[\,
   \frac{ I_{Pi_1Pi_2n_1n_2}(\omega-\veps_{Pi_1}) I_{n_1n_2k_1k_2}(\veps_{k_1}-\omega) }
        { \big( \omega - \veps_{n_1} + i\eta_{n_1}0 \big) \big( \bar{E}_{ik}^{(0)} - \omega - \veps_{n_2} + i\eta_{n_2}0 \big) } 
 + \{ 1 \leftrightarrow 2 \}
 \,\Bigg] \, .
\end{align}
We readily extract the desired three-electron contribution from Eq.~(\ref{eq:H2ik_ld_irr})
\begin{align}
\label{eq:3el}
\delta H_{ik}^{\rm 3el}&= 
-
\frac{1}{2} \, \sum_P (-1)^P \, \sum_{c}\sum_{n} \, 
\frac{1}{ \bar{E}_{ik}^{(0)} - \veps_{c} - \veps_{n} }
\Bigg[ \, 
I_{Pi_1Pi_2nc}(\bar{E}_{ik}^{(0)}-\veps_{c}-\veps_{Pi_1}) I_{nck_1k_2}(\veps_{k_1}+\veps_{c}-\bar{E}_{ik}^{(0)}) \nonumber\\
& \qquad\qquad\qquad\qquad\qquad\quad
+
I_{Pi_1Pi_2nc}(\veps_{c}-\veps_{Pi_2}) I_{nck_1k_2}(\veps_{k_2}-\veps_{c})
 + \{ 1 \leftrightarrow 2 \}
\, \Bigg] \, .
\end{align}
\end{widetext}
The total three-electron contribution can be obtained by studying the crossed diagram 
and the two-electron self-energy and vacuum-polarization graphs. 

By considering the one- and two-electron diagrams in the formalism with the redefined vacuum, we take into account all the necessary contributions describing the interaction between the $L$ and $K$ shells. In order to evaluate the total binding energies of Be-like ions, we have to add the QED contributions corresponding to the $1s^2$ core. This issue is discussed in details, e.g., in Refs.~\cite{Malyshev:2019:010501_R, Kozhedub:2019:062506}. 
As a result, our numerical approach rigorously takes into account all the contributions of the first and second orders of QED PT.
The electron-correlation contributions due to the exchange by three or more photons are accounted for within the Breit approximation in the present work. The corresponding calculations are based on the Dirac--Coulomb--Breit (DCB) Hamiltonian and performed by means of the large-scale configuration-interaction (CI) method in the basis of the Dirac--Sturm orbitals~\cite{Bratzev:1977:173, Tupitsyn:2003:022511, Kaygorodov:2019:032505}. The procedure how to merge the QED calculations with the higher-order interelectronic-interaction contributions in case of quasidegenerate levels was suggested first in Ref.~\cite{Malyshev:2019:010501_R} and described in more details in Ref.~\cite{Kozhedub:2019:062506}. Finally, we account for the nuclear recoil and nuclear polarization effects which lie beyond the external-field approximation, that is beyond the Furry picture. 



\begin{table}[t]
\centering

\renewcommand{\arraystretch}{1.1}

\caption{\label{tab:binding} 
         Binding energies (with the opposite sign) of the ground and singly excited states in Be-like xenon (in eV). 
         Comparison of the different approaches: A, B, C, and D. See the text for details.
         }
         
\resizebox{\columnwidth}{!}{%
\begin{tabular}{
                @{}c@{\quad}
                @{\,\,}l@{\,\,}
                @{\,\,}S[table-format=6.3]@{\,\,}
                @{\,\,}S[table-format=6.3]@{\,\,}
                @{\,\,}S[table-format=6.3]@{\,\,}
                @{\,\,}S[table-format=6.3]@{}
               }
               
\hline
\hline

   \multicolumn{1}{c}{\rule{0pt}{1.2em}$\Omega$~~~}    &
   \multicolumn{1}{c}{$V_{\rm eff}$~~~~}                &
   \multicolumn{1}{c}{A~~~~}                              &
   \multicolumn{1}{c}{B~~~~}                              &
   \multicolumn{1}{c}{C~~~~}                              &
   \multicolumn{1}{c}{D}                                  \\      
        
\hline   
                       
 \rule{0pt}{3.6ex} &  \multicolumn{5}{c}{$2s2s\, ^1S_0$} \\[0.75mm]
\cline{2-6}

\multirow{3}{*}{$1\times1$}  &  Coul   \rule{0pt}{3.6ex}                 &     101071.884  &     100970.193  &     100973.026  &          {---}   \\ 
                             &  CH                     &     101071.948  &     100973.924  &     100972.977  &     100973.569   \\ 
                             &  LDF                    &     101071.928  &     100973.451  &     100972.981  &     100973.400   \\[1.5mm] 

\multirow{3}{*}{$2\times2$}  &  Coul                   &     101071.884  &     100970.443  &     100973.244  &     100973.263   \\ 
                             &  CH                     &     101071.948  &     100974.157  &     100973.194  &     100973.246   \\ 
                             &  LDF                    &     101071.928  &     100973.682  &     100973.198  &     100973.237   \\[1.5mm] 

\multirow{3}{*}{$3\times3$}  &  Coul                   &     101071.884  &     100970.487  &     100973.278  &     100973.240   \\ 
                             &  CH                     &     101071.948  &     100974.199  &     100973.229  &     100973.241   \\ 
                             &  LDF                    &     101071.928  &     100973.724  &     100973.233  &     100973.236   \\[1.5mm] 

\hline
 \rule{0pt}{3.6ex} &  \multicolumn{5}{c}{$2s2p\, ^3P_0$} \\[0.75mm]
\cline{2-6}

\multirow{3}{*}{$1\times1$}  &  Coul   \rule{0pt}{3.6ex}                 &     100961.328  &     100866.432  &     100868.743  &     100868.704   \\ 
                             &  CH                     &     100961.373  &     100869.738  &     100868.699  &     100868.710   \\ 
                             &  LDF                    &     100961.356  &     100869.227  &     100868.703  &     100868.705   \\[1.5mm] 

\hline
 \rule{0pt}{3.6ex} &  \multicolumn{5}{c}{$2s2p\, ^3P_1$} \\[0.75mm]
\cline{2-6}

\multirow{3}{*}{$1\times1$}  &  Coul   \rule{0pt}{3.6ex}                 &     100938.533  &     100843.637  &     100845.987  &     100845.930   \\ 
                             &  CH                     &     100938.579  &     100846.944  &     100845.941  &     100845.935   \\ 
                             &  LDF                    &     100938.562  &     100846.433  &     100845.946  &     100845.930   \\[1.5mm] 

\multirow{3}{*}{$2\times2$}  &  Coul                   &     100938.533  &     100843.635  &     100845.975  &     100845.935   \\ 
                             &  CH                     &     100938.579  &     100846.941  &     100845.930  &     100845.941   \\ 
                             &  LDF                    &     100938.562  &     100846.430  &     100845.934  &     100845.935   \\[1.5mm] 

\hline
 \rule{0pt}{3.6ex} &  \multicolumn{5}{c}{$2s2p\, ^3P_2$} \\[0.75mm]
\cline{2-6}

\multirow{3}{*}{$1\times1$}  &  Coul   \rule{0pt}{3.6ex}                 &     100596.593  &     100501.446  &     100503.789  &     100503.754   \\ 
                             &  CH                     &     100596.656  &     100504.784  &     100503.744  &     100503.757   \\ 
                             &  LDF                    &     100596.637  &     100504.272  &     100503.748  &     100503.751   \\[1.5mm] 

\hline
 \rule{0pt}{3.6ex} &  \multicolumn{5}{c}{$2s2p\, ^1P_1$} \\[0.75mm]
\cline{2-6}

\multirow{3}{*}{$1\times1$}  &  Coul   \rule{0pt}{3.6ex}                 &     100533.196  &     100438.050  &     100440.465  &     100440.441   \\ 
                             &  CH                     &     100533.262  &     100441.390  &     100440.416  &     100440.445   \\ 
                             &  LDF                    &     100533.242  &     100440.878  &     100440.420  &     100440.439   \\[1.5mm] 

\multirow{3}{*}{$2\times2$}  &  Coul                   &     100533.197  &     100438.053  &     100440.477  &     100440.437   \\ 
                             &  CH                     &     100533.262  &     100441.392  &     100440.428  &     100440.439   \\ 
                             &  LDF                    &     100533.242  &     100440.880  &     100440.432  &     100440.434   \\[1.5mm]

\hline
\hline

\end{tabular}%
}

\end{table}

Let us now turn to the discussion of the numerical results. In Table~\ref{tab:binding}, we present the binding energies of the ground and singly excited states in Be-like xenon. The calculations are performed starting from the Coulomb potential as well as within the extended Furry picture. 
In the latter case, the core-Hartree (CH) and local Dirac--Fock (LDF) screening potentials are incorporated in Eq.~(\ref{eq:DirEq}). 
The description and applications of these potentials can be found, e.g., in Refs.~\cite{Malyshev:2017:022512, Sapirstein:2002:042501, Shabaev:2005:062105}. For the nuclear charge distribution, the Fermi model is used. The root-mean-square (rms) radius and the nuclear mass of the isotope~$^{132}{\rm Xe}$ are taken as in Ref.~\cite{Yerokhin:2015:033103}. 

As noted above, for the ground and $J=1$ states we construct the alternative perturbation series for both the single and the quasidegenerate levels. The type of PT is shown in the first column, where the size of the subspace~$\Omega$ is indicated. The columns labeled with A, B, C, and D demonstrate how the energies change when we successively take into account different contributions. In the column A, we present the values obtained within the Breit approximation by means of the CI method. From Table~\ref{tab:binding}, one can see that for the specific potential these results do not depend on the size of $\Omega$. We note that the $1 \times 1$ values are just the CI energies, whereas the $2 \times 2$ and $3 \times 3$ values are obtained as the eigenvalues of the matrices~$H$ which are constructed based on the CI calculations in accordance with the prescriptions from Ref.~\cite{Kozhedub:2019:062506}. 
The energies for the specific subspace~$\Omega$ in the column A vary slightly from the potential to potential. This variation is caused by the dependence of the positive-energy-states projectors in the DCB Hamiltonian on the initial approximation in our approach, see the discussion, e.g., in Refs.~\cite{Kaygorodov:2019:032505, Kozhedub:2019:062506}. 

The results in the column B are obtained by adding the first-order QED contributions, namely the self energy, vacuum polarization, and frequency-dependent correction of the one-photon exchange contribution, as well as the nuclear recoil and nuclear polarization contributions. We stress that for the quasidegenerate levels the inclusion of the terms is not quite additive because of the mixing. One can see that the results for different potentials demonstrate significant scatter. 
The scatter can be reduced by considering the second-order QED corrections. This is done in the column C, where we add the contributions of the two-electron self-energy and vacuum-polarization diagrams, the nontrivial QED part of the two-photon exchange contribution (beyond the Breit approximation), and the two-loop one-electron corrections. 
The difference between the calculations with the Coulomb, CH, and LDF potentials indeed decreases going from B to C. The values in the column C obtained for the screening potentials are shifted slightly with respect to the Coulomb ones. This results from a rearrangement of the perturbation series within the extended Furry picture.

From the column C, it is seen that the energy of the ground $2s2s \, ^1S_0$ state considerably shifts as we pass from $1 \times 1$ to $2 \times 2$. However, the value of this shift is almost independent on the initial approximation. The shift is explained by the accurate treatment of the mixing of the states within the model subspace. 
When we extend $\Omega$ by including the $(2p_{3/2}2p_{3/2})_0$ configuration, the ground-state energy acquires an additional shift which is smaller by an order of magnitude. 
The uncertainties of these calculations are determined, in particular, by the uncalculated screened QED contributions of the second order in $1/Z$. Nowadays, these corrections are inaccessible by the rigorous QED methods. We can estimate them approximately employing the model Lamb-shift (QEDMOD) operator which has been suggested recently in Refs.~\cite{Shabaev:2013:012513, Shabaev:2018:69} and successfully applied to the QED calculations in various atomic systems~\cite{Tupitsyn:2016:253001, Pasteka:2017:023002, Yerokhin:2017:042505:note, Machado:2018:032517, Kaygorodov:2019:032505, Zaytsev:2019:052504, Shabaev:2020:052502}. 
In order to estimate the screened QED effects of the second order in $1/Z$, we have introduced the QEDMOD operator into Eq.~(\ref{eq:DirEq}) and evaluated the two-photon exchange contribution in the Breit approximation using the related one-electron basis. The correction of interest was obtained by subtracting the corresponding contribution calculated without the QEDMOD operator.
The binding energies with these corrections included are shown in the column D. One can see that the $1 \times 1$ values for the $2s2s \, ^1S_0$ state tend to the eigenvalues of the matrices~$H$. 
However, the discrepancy of the results for two screening potentials is large. Within nondegenerate PT, the higher-order QED effects contribute significantly, and their contribution can not be neglected. On the other hand, in the column D the difference between the $2 \times 2$ and $3 \times 3$ values decreases, and the Coulomb results become closer to the ones obtained for the screening potentials. 
Based on all the results given in Table~\ref{tab:binding}, we conclude that the $3 \times 3$ subspace~$\Omega$ is sufficient for the proper treatment of the electron-correlation and QED effects on the ground-state binding energy. 
As for the $J=1$ states, the situation, in principal, is the same. However, the mixing is less pronounced than for the ground state. 

The values shown in the column D of Table~\ref{tab:binding} for the LDF potential and the maximum size of the subspace~$\Omega$ are employed as the final results. The uncertainties are obtained by summing quadratically several contributions. First, in addition to the numerical errors, we take into account the uncertainties of the nuclear-size effect and of the two-loop one-electron corrections~\cite{Yerokhin:2015:033103}. Second, we estimate the uncalculated higher-order QED contributions through several means. The QED corrections to the electron-correlation effects of third and higher orders are estimated according to the procedure from Ref.~\cite{Kozhedub:2019:062506}. The screening of the two-loop contributions is estimated by multiplying the corresponding term for $1s$ by the conservative factor $2/Z$. We take into account the scatter of the results obtained for the different potentials as well. 
Finally, having in mind that the calculations of the screened QED effects of the second order in $1/Z$ in the column D are approximate, we take the corresponding correction with the 100\% uncertainty. 
As a result, for the ground-state binding energy we obtain $E[2s2s \, ^1S_0]=-100973.236(44)$~eV, which deviates from $-100972.921(85)$~eV~\cite{Malyshev:2014:062517}. We have to admit that the higher-order QED effects have been underestimated in our previous calculations. We note that the current $1 \times 1$ value presented in the column C for the LDF potential, which is obtained within the approach closest to Ref.~\cite{Malyshev:2014:062517}, is also shifted with respect to the old one within the designated error bar. The reasons are in the different approach to evaluate the higher-order interelectronic-interaction contributions within the Breit approximation, the revised two-loop one-electron corrections, and the updated values of the fundamental constants~\cite{Mohr:2016:035009}. The binding energies of the singly excited levels are $E[2s2p \, ^3P_0]=-100\,868.705(42)$~eV, $E[2s2p \, ^3P_1]=-100\,845.935(42)$~eV, $E[2s2p \, ^3P_2]=-100\,503.751(42)$~eV, and $E[2s2p \, ^1P_1]=-100\,440.434(42)$~eV.
We consider the present procedure for the estimation of the theoretical uncertainties to be much more reliable that the one applied earlier in Ref.~\cite{Malyshev:2014:062517}, since it is based on the comprehensive analysis of the results obtained by using different starting potentials within alternative PT and on the direct evaluation of the higher-order QED effects via the QEDMOD operator.

\begin{table}[t]
\centering

\renewcommand{\arraystretch}{1.1}

\caption{\label{tab:transition} 
         Excitation energies in Xe$^{50+}$ (in eV).
         }
         
\begin{tabular}{S[table-format=3.4(1)]
                S[table-format=4.4(1)]
                S[table-format=4.4(1)]
                S[table-format=4.4(1)]
                c
               }
               
\hline
\hline

   \multicolumn{1}{c}{\rule{0pt}{1.0em} $2s2p \, ^3P_0$}    &
   \multicolumn{1}{c}{                  $2s2p \, ^3P_1$}    &
   \multicolumn{1}{c}{                  $2s2p \, ^3P_2$}    &
   \multicolumn{1}{c}{                  $2s2p \, ^1P_1$}    &
   \multicolumn{1}{c}{Ref.}                            \\     
        
\hline   

\multicolumn{4}{c}{Theory \rule{0pt}{1.0em}} \\   

\hline
                       
   104.531(9) &   127.300(9) &   469.484(7) &   532.802(7) &   \rule{0pt}{2.6ex}  TW   \\ 

 104.5(25)  &  127.3(25)  &  469.6(25)  &  532.9(2 5)  &  \cite{Kaygorodov:2019:032505}  \\
 
 104.475    &  127.282    &  469.449    &  532.877     &  \cite{Cheng:2008:052504}  \\
 
 104.663    &  127.475    &  470.004    &  533.401     &  \cite{Gu:2005:267}  \\ 
 
            &  127.168    &  469.25     &  532.62      &  \cite{Safronova:2000:1213}  \\
            
            &  127.301    &             &  532.854     &  \cite{Chen:1997:166}  \\
            
 104.482    &  127.267    &  469.386    &  532.759     &  \cite{Safronova:1996:4036}  \\
 
 103.722    &  126.846    &  468.338    &  532.766     &  \cite{Cheng:1979:111}  \\  
 
\hline

\multicolumn{4}{c}{Experiment \rule{0pt}{1.0em}} \\  

\hline
 
 \rule{0pt}{1.0em}   &  127.269(46)  &  469.474(81)  &  532.801(16)  &  \cite{Bernhardt:2015:144008}  \\
            
            &  127.260(26)  &               &               &  \cite{Feili:2005:48}  \\
            
            &  127.255(12)  &               &               &  \cite{Trabert:2003:042501}  \\
            
            
            
            

\hline
\hline

\end{tabular}%

\end{table}

The obtained binding energies are used to evaluate the excitation energies for the $2s2p \, ^{2S+1}P_{J}$ states presented in Table~\ref{tab:transition}. The uncertainties are estimated similarly to the binding energies. 
The major part of the one-electron contributions cancels when the binding energies are subtracted. This leads to a considerable reduction of the numerical error. Along with it, we take the same estimate of the higher-order QED correction to the electron-correlation contribution and the uncertainty related to the screened two-loop diagrams as for the ground state. We note that the total theoretical uncertainties of the excitation energies cover the results obtained for the different initial approximations, including the Coulomb potential, as well as the values with and without the QEDMOD correction given in the column D in Table~\ref{tab:binding}.
In Table~\ref{tab:transition}, we compare our excitation energies with the results of the previous relativistic calculations and recent measurements. 
In Ref.~\cite{Kaygorodov:2019:032505}, these energies were evaluated by means of the CI$+$QEDMOD method, and the uncertainties were estimated in a rather conservative way. However, the comparison with the results of the present work shows that the accuracy of this approach is at least one order of magnitude higher.
As for the experiments, our results are in perfect agreement with the most recent measurements performed in Ref.~\cite{Bernhardt:2015:144008}, especially for the $2s2p \, ^1P_1$ state, for which the experimental uncertainty is minimal. 
In Ref.~\cite{Bernhardt:2015:144008}, the excitation energies were measured for Be-like~$^{136}{\rm Xe}$ which has the different values of the nuclear mass and rms radius. The corresponding correction for the data given in Table II constitutes less than 1 meV and, for this reason, can be neglected.
On the other hand, the most precise measurement for the $2s2p \, ^3P_1$ state~\cite{Trabert:2003:042501} deviates from our result by almost 4 times the experimental uncertainty and by 5 times our theoretical uncertainty. The reason of this discrepancy is unclear to us. 
We note that within our approach the binding energies of both the $2s2p \, ^3P_1$ and $2s2p \, ^1P_1$ states are obtained in the framework of same PT, and the corresponding excitation energies are expected to be on the same level of accuracy.
Trying to shed some light on this issue, we have applied the developed \textit{ab initio} approach to evaluation of the excitation energy for the $2s2p \, ^3P_1$ state in Be-like molybdenum and uranium. The obtained values, $E_{\rm th}[{\rm Mo}]=90.005(4)$~eV and $E_{\rm th}[{\rm U}]=297.90(11)$~eV, are in agreement with the results of the available high-precision measurements: $E_{\rm exp}[{\rm Mo}]=89.983(20)$~eV~\cite{Denne:1989:3702} and $E_{\rm exp}[{\rm U}]=297.799(12)$~eV~\cite{Beiersdorfer:2005:233003}. In molybdenum, the electron--electron correlations are more important than in xenon, whereas in uranium the QED effects come to the fore. So, we can conclude that the observed discrepancy in xenon is hardly explained by the calculational problems with the correlation or QED effects. We hope that this discrepancy will trigger the new studies of Be-like systems. It should be stressed also that the achieved theoretical accuracy for Be-like uranium provides tests of QED 
as well as its two-loop part 
at the same accuracy level as in Li-like ions. There is also the ultra-precise measurement of the fine-structure~$^3P_1$--$^3P_1$ interval in Be-like argon~\cite{Draganic:2003:183001}. We will address this system in near future.



To summarize, \textit{ab initio} QED calculations of the binding energies of the ground and singly excited states in Be-like xenon have been performed with the most advanced methods available to date. The calculations merge the rigorous QED treatment up to the second-order contributions and the higher-order electron-correlation effects evaluated within the Breit approximation. For the first time, the ground state as well as the states with the total angular momentum equal to 1 are treated by means of perturbation theory for quasidegenerate levels. As a result, we have obtained the most precise theoretical predictions for the energy levels and $\Delta n =0$ intra-$L$-shell excitation energies, which are in perfect agreement with the most recent measurements~\cite{Bernhardt:2015:144008}. Meanwhile, some discrepancy with the previous experiment~\cite{Trabert:2003:042501} is found. New measurements with Be-like xenon and other Be-like ions are in demand. 



This work was supported by the grant of the President of the Russian Federation (Grant No. MK-1459.2020.2), and by RFBR and ROSATOM according to the research project No. 20-21-00098. A.V.M., M.Y.K., and V.M.S. acknowledge the support from the Foundation for the Advancement of Theoretical Physics and Mathematics ``BASIS''. D.A.G. acknowledges the support by RFBR (Grant No. 19-02-00974). The work of Y.S.K. and I.I.T. was supported by RFBR (Grant No. 18-03-01220).





\end{document}